\documentstyle[12pt]{article}

  \def\pp{{\mathchoice
          {
              \kern 1pt%
              \raise 1pt
              \vbox{\hrule width5pt height0.4pt depth0pt
                    \kern -2pt
                    \hbox{\kern 2.3pt
                          \vrule width0.4pt height6pt depth0pt
                          }
                    \kern -2pt
                    \hrule width5pt height0.4pt depth0pt}%
                    \kern 1pt
           }
            {
              \kern 1pt%
              \raise 1pt
              \vbox{\hrule width4.3pt height0.4pt depth0pt
                    \kern -1.8pt
                    \hbox{\kern 1.95pt
                          \vrule width0.4pt height5.4pt depth0pt
                          }
                    \kern -1.8pt
                    \hrule width4.3pt height0.4pt depth0pt}%
                    \kern 1pt
            }
            {
              \kern 0.5pt%
              \raise 1pt
              \vbox{\hrule width4.0pt height0.3pt depth0pt
                    \kern -1.9pt  
                    \hbox{\kern 1.85pt
                          \vrule width0.3pt height5.7pt depth0pt
                          }
                    \kern -1.9pt
                    \hrule width4.0pt height0.3pt depth0pt}%
                    \kern 0.5pt
            }
            {
              \kern 0.5pt%
              \raise 1pt
              \vbox{\hrule width3.6pt height0.3pt depth0pt
                    \kern -1.5pt
                    \hbox{\kern 1.65pt
                          \vrule width0.3pt height4.5pt depth0pt
                          }
                    \kern -1.5pt
                    \hrule width3.6pt height0.3pt depth0pt}%
                    \kern 0.5pt
            }
        }}

  \def\mm{{\mathchoice
                       {
                             \kern 1pt
               \raise 1pt    \vbox{\hrule width5pt height0.4pt depth0pt
                                  \kern 2pt
                                  \hrule width5pt height0.4pt depth0pt}
                             \kern 1pt}
                       {
                            \kern 1pt
               \raise 1pt \vbox{\hrule width4.3pt height0.4pt depth0pt
                                  \kern 1.8pt
                                  \hrule width4.3pt height0.4pt depth0pt}
                             \kern 1pt}
                       {
                            \kern 0.5pt
               \raise 1pt
                            \vbox{\hrule width4.0pt height0.3pt depth0pt
                                  \kern 1.9pt
                                  \hrule width4.0pt height0.3pt depth0pt}
                            \kern 1pt}
                       {
                           \kern 0.5pt
             \raise 1pt  \vbox{\hrule width3.6pt height0.3pt depth0pt
                                  \kern 1.5pt
                                  \hrule width3.6pt height0.3pt depth0pt}
                           \kern 0.5pt}
                       }}

\catcode`@=11
\def\un#1{\relax\ifmmode\@@underline#1\else
        $\@@underline{\hbox{#1}}$\relax\fi}
\catcode`@=12

\let\du=\du                     

\def\a{\alpha}
\def\b{\beta}
\def\c{\chi}

\def\f{\phi}

\def\j{\psi}
\def\k{\kappa}
\def\l{\lambda}
\def\m{\mu}
\def\n{\nu}

\def\p{\pi}
\def\q{\theta}

\def\s{\sigma}
\def\t{\tau}

\def\D{\Delta}
\def\F{\Phi}

\def\L{\Lambda}
\def\O{\Omega}

\def\ve{\varepsilon}

\def\vq{\vartheta}

\def\cm{{\cal M}}

\def\co{{\cal O}}

\def\cy{{\cal Y}}

\def\bo{{\raise-.5ex\hbox{\large$\Box$}}}               
\def\pa{\partial}                                       
\def\pr{\prod}                                          
\def\TH{{\raise.2ex\hbox{$\displaystyle \bigodot$}\mskip-4.7mu \llap H \;}}
\def\face{{\raise.2ex\hbox{$\displaystyle \bigodot$}\mskip-2.2mu \llap {$\ddot
        \smile$}}}                                      

\def\sp#1{{}^{#1}}                              
\def\VEV#1{\left\langle #1\right\rangle}        
\def\abs#1{\left| #1\right|}                    
\def\leftrightarrowfill{$\mathsurround=0pt \mathord\leftarrow \mkern-6mu
        \cleaders\hbox{$\mkern-2mu \mathord- \mkern-2mu$}\hfill
        \mkern-6mu \mathord\rightarrow$}
\def\dvec#1{\vbox{\ialign{##\crcr
        \leftrightarrowfill\crcr\noalign{\kern-1pt\nointerlineskip}
        $\hfil\displaystyle{#1}\hfil$\crcr}}}           
\def\dt#1{{\buildrel {\hbox{\LARGE .}} \over {#1}}}     

\def\frac#1#2{{\textstyle{#1\over\vphantom2\smash{\raise.20ex
        \hbox{$\scriptstyle{#2}$}}}}}                   
\def\sfrac#1#2{{\vphantom1\smash{\lower.5ex\hbox{\small$#1$}}\over
        \vphantom1\smash{\raise.4ex\hbox{\small$#2$}}}} 
\def\bfrac#1#2{{\vphantom1\smash{\lower.5ex\hbox{$#1$}}\over
        \vphantom1\smash{\raise.3ex\hbox{$#2$}}}}       
\def\afrac#1#2{{\vphantom1\smash{\lower.5ex\hbox{$#1$}}\over#2}}    

\def\[{\lfloor{\hskip 0.35pt}\!\!\!\lceil}
\def\]{\rfloor{\hskip 0.35pt}\!\!\!\rceil}
\def\Lag{{\cal L}}
\def\du#1#2{_{#1}{}^{#2}}

\def\fracm#1#2{\hbox{\large{${\frac{{#1}}{{#2}}}$}}}
\def\ha{{\fracmm12}}

\def\un{\underline}
\def\fracmm#1#2{{{#1}\over{#2}}}

\def\low#1{{\raise -3pt\hbox{${\hskip 0.75pt}\!_{#1}$}}}

\def\Dot#1{\buildrel{_{_{\hskip 0.01in}\bullet}}\over{#1}}
\def\dt#1{\Dot{#1}}

\newskip\humongous \humongous=0pt plus 1000pt minus 1000pt
\def\caja{\mathsurround=0pt}
\def\eqalign#1{\,\vcenter{\openup2\jot \caja
        \ialign{\strut \hfil$\displaystyle{##}$&$
        \displaystyle{{}##}$\hfil\crcr#1\crcr}}\,}
\newif\ifdtup

\def\ref#1{$\sp{#1)}$}

\def\pl#1#2#3{Phys.~Lett.~{\bf {#1}B} (19{#2}) #3}
\def\np#1#2#3{Nucl.~Phys.~{\bf B{#1}} (19{#2}) #3}
\def\prl#1#2#3{Phys.~Rev.~Lett.~{\bf #1} (19{#2}) #3}
\def\pr#1#2#3{Phys.~Rev.~{\bf D{#1}} (19{#2}) #3}
\def\cqg#1#2#3{Class.~and Quantum Grav.~{\bf {#1}} (19{#2}) #3}

\def\ijmp#1#2#3{Int.~J.~Mod.~Phys.~{\bf A{#1}} (19{#2}) #3}

\def\ibid#1#2#3{{\it ibid.}~{\bf {#1}} (19{#2}) #3}

\parskip=\medskipamount                 
\thicklines                         

\begin{document}
\thispagestyle{empty}

{\hbox to\hsize{
\vbox{\noindent KL~--~TH 00/06   \hfill December 2000 \\
hep-th/0009187 \hfill                   revised version }}}

\noindent
\vskip1.3cm
\begin{center}

{\Large\bf Exact Renormalization Flow and \vglue.1in
            Domain Walls from Holography }
\vglue.3in

Sergei V. Ketov 
\footnote{Supported in part by the `Deutsche Forschungsgemeinschaft'}

{\it Fachbereich Physik\\
     Universit\"at Kaiserslautern\\
     Erwin Schr\"odinger Strasse \\
     67653 Kaiserslautern, Germany}
\vglue.1in
{\sl ketov@physik.uni-kl.de}
\end{center}
\vglue.2in
\begin{center}
{\Large\bf Abstract}
\end{center}

The holographic correspondence between 2d, N=2 quantum field theories and 
classical 4d, N=2 supergravity coupled to hypermultiplet matter is proposed. 
The geometrical constraints on the target space of the 4d, N=2 non-linear 
sigma-models in N=2 supergravity background are interpreted as the exact 
renormalization group flow equations in two dimensions. Our geometrical 
description of the renormalization flow is manifestly covariant under general
reparametrization of the 2d coupling constants. An explicit exact solution to 
the 2d renormalization flow,  based on its dual holographic description in 
terms of the Zamolodchikov metric, is considered in the  particular case of 
the four-dimensional NLSM target space described by the $SU(2)$-invariant 
(Weyl) anti-self-dual Einstein metrics. The exact regular (Tod-Hitchin) 
solutions to these metrics are governed by the Painlev\'e VI equation, and 
describe domain walls. 

\newpage

\section{Introduction}

A holographic correspondence was first proposed by 't Hooft \cite{th} and 
Susskind \cite{suss} in the particular context of black hole physics. 
According to the original formulation \cite{th,suss}, information on degrees
 of freedom inside a volume can be encoded in a surface enclosing this 
volume. There are many reasons to believe that the holographic principle may
 be valid far beyond its original framework, though a precise formulation of
 this principle is still lacking. The essence of the contemporary use of the 
 holographic principle amounts to the assertion that a classical field 
theory (with gravity) in a volume is equivalent to certain Quantum Field 
Theory (QFT) (without gravity) defined on the boundary of this volume. This
idea may be elevated to the existence of an equivalent description of QFT in
 terms of classical gravity in higher dimensions, or it may also be 
considered as the manifestation of the fundamental equivalence between 
quantum gauge field theories and strings \cite{pol}. For example, the very 
popular AdS/CFT correspondence in its original formulation 
\cite{ads} relates the type-IIB superstring theory on $AdS_5\times S^5$ with 
the {\it four-dimensional} (4d), N=4 supersymmetric $SU(N)$ Yang-Mills theory
 on the boundary of the $AdS_5$ space. The string loop corrections are 
proportional to $N^{-2}$, whereas the $\a'$-corrections are proportional to 
$\l^{-1/2}$, where $\l=g^2_{\rm YM}N$ is the `t Hooft coupling, so that the 
large-N and large-$\l$ (strong coupling) limit can be investigated in the 
five-dimensional AdS-supergravity approximation (see ref.~\cite{adsr} for a 
review). 

The Maldacena conjecture \cite{ads} relates two field theories in different 
dimensions, both having very high amount of symmetry (e.g., conformal 
symmetry, extended supersymmetry, electric-magnetic self-duality). 
Nevertheless, the holographic correspondence appears to be valid even if 
many of these symmetries are broken. For example, a massive deformation of 
the N=4 SYM theory breaks down both N=4 supersymmetry and conformal 
invariance, which result in a non-trivial Renormalization Group (RG) flow. 
The radial coordinate of the $AdS_5$ gives the natural scale to the 4d 
quantum gauge theory. The holographic duality in this context means the 
identification of the (classical) five-dimensional supergravity equations 
of motion in the bulk with the (quantum) RG-flow equations in the dual 
(large-$N$) quantum gauge theory on the four-dimensional boundary 
\cite{flow}. A specific example of the RG flow from the N=4 (superconformal)
 Yang-Mills theory in the UV to an N=1 (superconformal) gauge theory in the 
IR was given in ref.~\cite{usc}, where this flow was identified with a 
domain wall (BPS) solution to the five-dimensional (gauged) N=8 supergravity
 connecting two $AdS_5$ vacua. Away from a few known and highly symmetric 
examples of the holographic correspondence, it is far from being clear 
{\it why} does the holography exist, and, perhaps, most importantly, 
{\it where} does the holographic principle apply.

To test the holographic principle and determine the area of its 
applicability, it is worthy to investigate more explicit examples of the
holographic correspondence under the circumstances that are different from 
the standard AdS/CFT, e.g. in lower dimensions where many deep theorems 
about generic QFT and gravity are available, thus leaving less room for 
speculations. So we replace the 4d quantum Yang-Mills theories, playing 
the key role in the AdS/CFT correspondence, by the 2d quantum Non-Linear 
Sigma-Models (NLSM) with torsion and scalar potential. The 2d NLSM are known
 to share many key features with the 4d gauge theories, such as 
conformal invariance, renormalizabiliy, solitons, asymptotic freedom, etc. 
\cite{nlsmbook}. The similarity between the RG flows in 4d gauge theories 
and the RG flows in 2d NLSM, in the context of the AdS/CFT correspondence, 
was also noticed in ref.~\cite{schm}. To get control over quantum 
non-perturbative issues, we also impose N=2 extended supersymmetry in the 2d
 QFT under consideration, and require its integrability  (i.e. the inifinitely
 many conservation laws, or a factorizability of the S-matrix).
We find that the RG flow in the 2d, N=2 QFT admits the most natural 
description in terms of the effective {\it four-dimensional} N=2 
supergravity coupled to hypermultiplet matter. The hypermultiplet scalars 
represent the couplings of the 2d, N=2 QFT under consideration, whereas the 
four-dimensional N=2 supergravity serves as the geometrical background for 
the hypermultiplet NLSM (or as the source for the stress-energy tensor), as 
usual. The metric of the 4d, N=2 NLSM is identified with the Zamolodchickov 
metric of the 2d, N=2 QFT. This correspondence is nothing but the off-critical
 extension of the well-known world-sheet/spacetime correspondence 
\cite{polbook} between the 2d, N=2 superconformal field theories and the 
effective four-dimensional N=2 supergravity theories, which is known in the 
type-II superstring compactification on Calabi-Yau spaces, under the 
preservation of integrability and supersymmetry. We find that the RG flow 
in the 2d, N=2 QFT can be naturally described by the radial dependence of 
the NLSM metric {\it solution}. The NLSM dual description automatically 
accommodates the general coordinate invariance in the space of QFT couplings 
\cite{hol}, since it amounts to the reparametrizational invariance of the NLSM 
target space, while the existence of the natural radial coordinate in the 
NLSM target space is guaranteed by the quaternionic nature of the NLSM 
metric. Taken together, this means the existence of a holographic 
correspondence between certain two-dimensional (2d) N=2 supersymmetric QFT and 
classical four-dimensional (4d) N=2 supergravity with hypermultiplet matter. 

Yet another simplification of our setup, in comparison to the 
higher-dimensional AdS/CFT correspondence, is the absence of a non-trivial 
four-form flux (or RR background) that must be present in the $AdS_5$ space 
where it results in the gauging of an abelian isometry of the hypermultiplet 
NLSM --- see ref.~\cite{ovr} where the domain-wall solutions from the M-theory 
compactification on Calabi-Yau threefolds were discussed. The use of holography
 in application to the 2d QFT versus 4d QFT apparently leads to the effective 
description in terms of the 4d supergavity versus 5d supergravity, 
respectively.

As an explicit example, we examine in detail the simplest non-trivial case 
of a single matter hypermultiplet in 4d, N=2 supergravity background from the
viewpoint of holography. The Zamolodchikov metric then appears to be a 
four-dimensional Einstein-Weyl metric, whose {\it exact} regular solutions 
are given by Tod-Hitchin metrics \cite{tod,hit}. We demonstrate that the 
Einstein-Weyl gravity equations can be interpreted as the RG flow equations,
 whereas their regular solutions describe the domain walls relating two 
(UV and IR) fixed points. A derivation of any exact RG domain wall solution 
is known to be a formidable problem in physics. To the best of our 
knowledge, the Tod-Hitchin metrics were never considered in the context of the
 holographic correspondence, while they also give the remarkable 
connection between the RG flow in the integrable 2d, N=2 QFT and the 
standard integrable non-linear equations of mathematical physics, such as the 
Painlev\'e VI equation.

The paper is organized as follows. In sect.~2 generic 2d CFT and QFT are
discussed from the viewpoint of the holographic correspondence. In sect.~3 we 
review the special features of this correspondence after adding N=2 
supersymmetry. Sect.~4 is devoted to the covariant differential equations on 
the Zamolodchikov metric and their homegeneous solutions in the case of the
2d, N=2 superconformal field theories whose 4d, N=2 supergravity duals are 
described in terms of a single hypermultiplet. The relation between the Weyl
self-duality and RG flow in this particular case is given in sect.~5. The 
explicit exact solutions to the RG flow, based on the regular 
Tod-Hitchin metrics, are given in sect.~6 where their relation to the
Painlev\'e VI equation is explained. Sect.~7 is our conclusion. Basic facts
about theta functions are summarized in Appendix.

\section{QFT and Zamolodchikov theorems in 2d}

Let's consider an abstract {\it two-dimensional} (2d) Conformal Field Theory 
(CFT)  of central charge $c$. Let $\Lag_{\rm CFT}$ be its Lagrangian. A 2d 
QFT is defined by the perturbed Lagrangian
$$ \Lag_{\rm QFT}= \Lag_{\rm CFT} +\sum_i\l_i\co_i~,\eqno(2.1)$$
where $\co_i$ are some local (normally ordered) composite operators, and
$\l_i$ are the associated (finite, not infinitesimal) coupling constants. We
always assume that the 2d QFT (2.1) is renormalizable; in this case the 
perturbations $\co_i$ are also called renormalizable.

The 2d QFT (2.1) may define another 2d CFT provided the operators $\co_i$ are 
chosen to be of conformal weights (1,1). In this case, the associated 
perturbations are called marginal, while they are particularly
relevant for string theory \cite{polbook}. In fact, all 2d CFT can be described
by the use of marginal perturbations (see, e.g., ref.~\cite{cftbook} for 
a review), so that a choice of the CFT to begin with is at our disposal. 
Though a generic CFT does not have a convenient Lagrangian description 
\cite{cftbook}, the details of its Lagrangian are not going to be relevant for
our purposes. As is well known, the 2d CFT are most efficiently described by 
the use of their symmetries. The initial CFT can be chosen to be maximally 
symmetric. Any 2d CFT defines an integrable field theory in the sense that it 
has the infinitely many conservation laws. This property can be naturally 
generalized: a 2d QFT is called integrable provided that it possesses the
infinitely many conservation laws or, equivalently, if it has a factorizable 
S-matrix \cite{zam1}. The Zamolodchikov techniques of derivation of the 
integrable (massive) deformations of CFT are essentially based on demanding 
the infinite number of conserved currents to survive the perturbations away 
from the criticality \cite{zam1}.

The fundamental difference between CFT and QFT is due to the fact that the 
latter has an (energy) scale $\m$, whereas the former is scale invariant by
definition. It is therefore natural to introduce the (one-parameter) family of
 QFT related to each other by a change of scale. Changing the scale in the 
parameter space $\cm=\{\l_i\}$ defines a `flow' known as the RG flow. The 
fixed points of this flow are CFT's. A QFT is characterized by a point in 
$\cm$. The RG trajectory through this point allows us to define the UV and 
 IR limits of a given QFT, by taking $\m\to\infty$ and $\m\to 0$, 
respectively. The CFT we started with can then be naturally identified with 
the UV fixed point of QFT. We identify the scale $\m$ of 2d QFT with one of
its coupling constants $\l_0\,$.

As regards the IR-limit of a QFT, it may be either a massive (non-conformal) 
field theory or yet another CFT \cite{zam2}. In the latter case, the RG flow 
interpolates between the UV- and IR-fixed points, so that it can be 
interpreted as a {\it domain wall} in $\cm$. This is precisely the case that 
we would like to investigate in this paper, from the holographic point of view.
 Since we are interested in 2d QFT's, we have the advantage of having two 
very powerful tools due to Zamolodchikov \cite{zam2,zam3}. The famous
Zamolodchikov theorems imply (i) the existence of a metric in $\cm$, which 
is defined by the two-point function of perturbing opertators on the plane at 
a fixed distance,
$$ G_{ij}=\VEV{\co_i\co_j}~,\eqno(2.2)$$
and (ii) the existence of a function (called $c$-function) in $\cm$, which 
monotonically decreases along the RG flow,
$$ \dt{c}\,\equiv\fracmm{dc}{dt}\leq 0~,\qquad c=c(t)~,\quad t=\ln\m~,
\eqno(2.3)$$
and whose fixed points $(\dt{c}\,=0)$ correspond to the RG-fixed points, i.e. 
the CFT's. The RG beta-functions in QFT are defined by~\footnote{A summation
over repeated indices is always assumed.}
$$ \dt{\l}_i\,=\b_i(\l)=\D_{ij}\l_j+C_{ijk}\l_j\l_k+\ldots,~¸
\eqno(2.4)$$
where $\D_{ij}$ represent the anomalous scaling dimensions, $C_{ijk}$ 
are the Operator Product Expansion (OPE) coefficients of the operators 
$\vec{\co}$ in 2d QFT, and the dots stand for the higher order terms 
(in $\vec{\l}$) of the expansion of the beta-functions in power series near 
a fixed point. The OPE coefficients $C_{ijk}$ are universal provided that 
$\D_{ij}=0$. 

The $c$-function of Zamolodchikov can be considered as a function of the 
coupling constants $\l_i$, whose stationarity implies criticality, i.e.
$$ \fracmm{\pa c}{\pa \l_i}=0\quad{\rm is~equivalent~to} \quad \b_i=0~,
\eqno(2.5)$$
while the critical value of the $c$-function at a fixed point can be 
identified with the central charge $c$ of the corresponding CFT,
$$ c(\l_{\rm crit.})=c~,\quad \b_i(\l_{\rm crit.})=0~.\eqno(2.6)$$
It should be stressed that eq.~(2.5) is merely an on-shell relation. 
A stronger off-shell conjecture in the form
$$  \fracmm{\pa c}{\pa \l_i}=K_{ij}\b_j \eqno(2.7)$$
with some invertible matrix  $K_{ij}$ often appears in the literature
(together with yet another proposal that $K_{ij}$ may even be proportional to
Zamolodchikov metric $G_{ij}$), despite of the fact that explicit multi-loop 
calculations in the 2d NLSM with torsion do not support eq.~(2.7) --- 
see refs.~\cite{jj,nlsmbook} for details.

In string theory, the massless physical modes (associated with a spacetime 
metric $g_{\m\n}$, an antisymmetric tensor $b_{\m\n}$, and a dilaton $\F$) of
 a string are described by marginal deformations of 2d CFT, while the 
low-energy string effective action in spacetime is determined by the vanishing
 RG beta-functions of the 2d NLSM describing string propagation in the 
background of its massless modes \cite{polbook,nlsmbook}. In this context, 
Zamolodchikov's $c$-theorem just guarantees the existence of the string 
effective action. In fact, since the Zamolodchikov $c$-function also makes 
sense off criticality, it is always possible to promote the coupling constants
 $\l_i$ to the scalar fields $\f_i(x)$ in spacetime $(x^{\m})$, whose 
low-energy effective action is given by the NLSM with Zamolodchikov metric 
$G_{ij}(\f)$, and whose vacuum expectation values are $\VEV{\f_i}=\l_i$. The 
scalar fields $\f_i$ then develop a non-trivial scalar potential 
$V(\f)$ in the effective Lagrangian. The RG flow interpolates between
 two different extrema of $V(\f)$ --- here is the name `domain wall' comes 
from ({\it cf.} refs.~\cite{flow,usc}),
$$ \fracmm{\pa V}{\pa \f_i}=0\quad\leftrightarrow\quad\dt{c}\,=0~.\eqno(2.8)$$
We are thus led to the following (bosonic) low-energy 
effective  Lagrangian \cite{hol}:
$$ \Lag(\f,g) = \fracm{1}{2}G_{ij}(\f)\pa^{\m}\f^i\pa_{\m}\f^j -V(\f) 
 -  \fracm{1}{2}R~,\eqno(2.9)$$
which is nothing but the minimal coupling of the NLSM (with the NLSM target 
space metric equal to the Zamolodchikov metric $G$, and the scalar potential 
$V$) to background gravity $g_{\m\n}$.~\footnote{The dimensional NLSM coupling 
constant in front of the NLSM action and the gravitational \newline ${~~~~~}$
constant in front of the Einstein-Hilbert action are both set to be one in 
eq.~(2.9).} From the 2d perspective, the spacetime metric $g_{\m\n}$ in the 
effective Lagrangian (2.9) represents marginal deformations, while the NLSM 
scalars $\f_i$ appears as the sources for the local operators $\co_i$. Given 
the energy scale $\m$ that is much smaller than the cut-off scale 
$\m_c\sim e^{kr}$ of QFT, the Lagrangian (2.9) can be trusted. The `low-energy'
approximation means that we are only interested in the local part of the QFT
effective action, or the two-point correlators of eq.~(2.2). As was 
demonstrated in ref.~\cite{hol}, in the context of the AdS/CFT correspondence,
 the identification of the five-dimensional low-energy effective Lagrangian 
with eq.~(2.9) indeed leads to the standard Callan-Symanzik RG equations for 
the holograhphic RG flow in the dual four-dimensional QFT. At very low 
energies, where merely the potential term in eq.~(2.9) is relevant, its 
IR value determines the 4d central charge $c$ (i.e. the holographic Weyl 
anomaly, or the critical exponent $k$) of the CFT on the $AdS_5$ boundary, 
$\lim_{\m\to 0}V=\fracm{3}{4}k^{-2}=\fracm{3}{16}c^{-2/3}$ \cite{hs}. We 
would like to describe the RG evolution in 2d QFT by identifying its parameter
 space $\cm$ with the NLSM target space in eq.~(2.9). A derivation of the 
exact Zamolodchikov metric (not just in the vicinity of a fixed point) is the 
much more complicated problem than a calculaton of the central charge, and, 
to our knowledge, in the context of the holographic correspondence, it was 
 never addressed elsewhere.
 
\section{Adding N=2 supersymmetry}

The holographic correspondence formulated in the preceeding section is very 
general, but it cannot be made more specific unless we add more structure to
the dual QFT/gravity pairs. N=2 extended supersymmetry (eight supercharges) is
 the natural symmetry that puts both QFT and gravity under control 
(the `N=2 wonderland'). It is worth mentioning here that the 5d, N=1 AdS
superalgebra $SU(2,2|1)$ corresponds to N=2 supersymmetry in 4d, or N=(2,2) 
superconformal symmetry in 2d. The supersymmetric world-sheet/spacetime
correspondence is well known in the standard superstring compactification, with
 the 2d (world-sheet) N=2 superconformal models being used as the building 
blocks of the spacetime supersymmetric superstring vacua \cite{polbook}. 
For example, as regards the closed type-II superstrings (compactified on a 
Calabi-Yau space $\cy$), their low-energy effective action is given by 
the four-dimensional N=2 supergravity theory, 
whose matter couplings have the `special' K\"ahler geometry in the sector 
of N=2 vector  multiplets \cite{vec} and the quaternionic-K\"ahler geometry 
in the sector of hypermultiplets \cite{bw}. We are going to concentrate here 
on hypermultiplets, each containing four real scalars and a Dirac hyperino. 
The corresponding (unique) action is given by the N=2 (locally) supersymmetric
 extension of eq.~(2.9) in four spacetime dimensions.~\footnote{The number of 
spacetime dimensions could also be five or six, as long as it does not affect 
the \newline ${~~~~~}$ NLSM target space.} In the context of the superstring 
compactification, the Zamolodchikov metric $G_{ij}$ of the underlying 2d, N=2 
Superconformal Field Theory (SCFT) is identified with the 
$4(h_{1,2}+1)$-dimensional quaternionic metric on the moduli space $\cm$ of 
the Calabi-Yau threefold $\cy$, whose Hodge number $h_{1,2}$ is just the 
number of the harmonic $(1,2)$-forms on the threefold $\cy$. Taken together, 
these facts amount to the existence of the holographic correspondence between 
the 2d, N=2 SCFT and the 4d, N=2 supergravities arising as the low-energy 
effective field theories of type-II superstrings.

This N=2 CFT/supergravity correspondence can be extended to the off-critical 
holographic correspondence between N=2 supersymmetric QFT and certain 4d, N=2 
supergravities, since the Zamolodchikov metric still makes sense off 
criticality. In the early nineties, Cecotti and Vafa \cite{cv} applied 
topological methods to study integrable (massive) deformations of 2d, N=2 CFT
(see ref.~\cite{fmvw} too). In particular, the effective NLSM metrics for some 
2d, N=2 CFT and QFT were calculated by identifying the Zamolodchikov metric 
with the metric of the N=2 supersymmetric ground states. The ground state 
metric, as the function  of perturbation parameters $\vec{\l}$, in the 
integrable case obeys the classical Toda (or affine Toda)  equations, which 
arise as the flatness conditions for certain holomorphic and anti-holomorphic 
connections in the vacuum bundle over $\cm$ \cite{cv,fmvw}. The coefficients 
 of the Toda-like integrable equations are just given by the topological 
correlation functions \cite{cv}. In the critical (N=2 SCFT) case, the exact 
solutions to these equations are governed by some holomorphic data of moduli, 
whereas  the underlying data in the case of an integrable N=2 QFT is not 
holomorphic. Unfortunately,  all examples of the massive (integrable) 
deformations of the N=2 supersymmetric Landau-Ginzburg models, considered in 
refs.~\cite{cv,fmvw}, lead to the massive field theories in the infra-red 
limit. In particular, the Cecotti-Vafa solutions associated with {\it generic}
  deformations of N=2 SCFT, are not the domain-wall solutions. The letter are
 associated with the special deformations whose choice is not apparent in 2d.
 The use of holography and R-invariance makes it possible to identify the 
relevant deformations in the dual picture and, sometimes, explicitly derive 
the corresponding regular Zamolodchikov metrics describing the domain-walls 
preserving both N=2 supersymmetry and R-symmetry (sects.~5 and 6). 

It is worth mentioning that the presence of a non-trivial scalar potential in
the N=2 supersymmetric extension of eq.~(2.9) is not in conflict with N=2
supersymmetry, which prohibits a superpotential in renormalizable 4d, N=2 QFT.
 In fact, any non-trivial N=2 NLSM kinetic terms with non-vanishing central 
charges give rise to a (unique) non-trivial scalar potential --- see, e.g., 
ref.~\cite{ku} for some explicit examples.~\footnote{The 4d NLSM are 
non-renormalizable.} Gauging the NLSM isometries also gives rise to a 
non-trivial scalar potential. We didn't attempt to calculate the scalar 
potential explicitly.

\section{Universal hypermultiplet and Wolf spaces} 

As is well known, the N=2 scalar (hypermultiplet) couplings in the 
four-dimensional N=2 supergravity are described by the NLSM with 
quaternionic-K\"ahler target spaces of negative scalar curvature \cite{bw}.
 In our context (sects.~2 and 3) this result implies that the
 Zamolodchikov metric in a 2d, N=2 QFT is also quaternionic-K\"ahler. Moreover,
 any Einstein space of a constant negative scalar curvature admits a natural 
radial coordinate \cite{feld}, whose existence is the central element of the 
holographic correspondence. 

We shall only consider in detail the simplest non-trivial case of a 
single hypermultiplet, which corresponds to four coupling constants. 
One of the coupling constants is identified with the RG parameter $\m$, 
so that we are going to deal with the RG flow of three couplings. In the
four-dimensional type-II superstring models, the dilaton belongs to 
a hypermultiplet, together with an axion and a complex RR scalar. In the 
literature this hypermultiplet is called the universal hypermultiplet 
\cite{univ}, even though there seem to be nothing `universal' in its coupling 
to either N=2 supergravity or N=2 matter multiplets, from the viewpoint of N=2 
supersymmetry, when compared to other hypermultiplets.

All {\it homogeneous} quaternionic-K\"ahler spaces are  classified 
\cite{alex}, while they are naturally associated with the 2d, N=2 or N=4 SCFT 
via the standard Kazama-Suzuki construction \cite{cftbook}. The {\it 
four-dimensional} target space of the 4d, N=2 NLSM in N=2 supergravity 
background is (Weyl) Anti-Self-Dual (ASD) and Einstein (see ref.~\cite{besse}
for mathematical details). The ASD Einstein metric should, therefore, obey the
 differential equations  
$$ W^+_{abcd}=0 \quad{\rm and}\quad  R_{ab}=\ha\L g_{ab}~,\qquad \L=-24\k^2~,
\eqno(4.1)$$
where $W=W^- + W^+$ is the Weyl tensor, $R_{ab}$ is the Ricci 
tensor, $a,b,c,d=1,2,3,4$, and $\k$ is the gravitational coupling constant.
It is worth mentioning that eqs.~(4.1) are manifestly covariant under general
 reparametrizations of the coupling constants. In other words, the general 
coordinate invariance of the RG flow in the space of couplings is guaranteed
in the dual (NLSM) description.

Given a simple Lie group $G$, the associated quaternionic {\it symmetric} 
 (homogeneous) space is unique; it is called the Wolf space \cite{wolf},
$$ \fracmm{G}{H_{\bot}\times SU(2)_{\c}}~~~,\eqno(4.2)$$
where $\c$ is the highest root of $G$,  the $SU(2)_{\c}$ is the subalgebra of
 $G$, associated with the root $\j$, and $H_{\bot}$ is the centralizer of 
 $SU(2)_{\c}$ in $G$. There are only two four-dimensional Wolf cosets of
negative scalar curvature, $SO(4,1)/SO(4)$ and $SU(2,1)/U(2)$. Both Wolf
spaces have the $SU(2)$ isometry that can be identified with the 
automorphisms (R-symmetry) $SU(2)$ of the rigid N=2 supersymmetry algebra.

A generic metric, possessing the $SU(2)$ isometry, is most conveniently
described (like in general relativity) by the Bianchi IX formalism where the 
$SU(2)$ symmetry is manifest. Given a `radial' coordinate $r$  and 
`Euler angles' $(\q,\j,\f)$, one introduces the $SU(2)$-covariant one-forms, 
$$\eqalign{
\s_1~=~&+\fracm{1}{2}(\sin\j d\q -\sin\q\cos\j d\f)~, \cr 
\s_2~=~&-\fracm{1}{2}(\cos\j d\q +\sin\q\sin\j d\f)~,\cr
\s_3~=~&+\fracm{1}{2}(d\j+\cos\q d\f)~,\cr}\eqno(4.3)$$
which satisfy the relation $\s_i\wedge \s_j=\ha \ve_{ijk}d\s_k$. The standard
metric, associated with the symmetric (Euclidean $AdS_4$) space 
$SO(4,1)/SO(4)$, is conformally flat, 
$$ ds^2 =\fracmm{1}{(1-r^2)^2}\left[ dr^2 + r^2(\s^2_1+\s^2_2+\s^2_3)
\right]~.\eqno(4.4)$$
The boundary of $AdS_4$ at $(r\to 1^-)$ is given by a three-dimensional 
sphere $S^3$, while the four-dimensional conformal structure in the ball 
$r^2<1$ induces the conformal structure on $S^3$. Unfortunately, the 
non-topological (non-trivial) QFT in three dimensions are not conformally 
invariant, and a conformal anomaly does not exist in three dimensions (like in
 any other odd-dimensional space). Therefore, we should not expect the 
existence of a non-trivial IR-fixed point, when starting from the $AdS_4$ 
and applying the RG flow.

The remaining symmetric Wolf space $SU(2,1)/U(2)$ is perfectly suitable for our
purposes. The  natural metric in this space is given by the so-called Bergmann
 metric which is dual to the standard Fubini-Study metric \cite{besse},
$$ ds^2 =\fracmm{dr^2}{(1-r^2)^2}+\fracmm{r^2}{(1-r^2)^2}\s^2_2
+\fracmm{r^2}{(1-r^2)}(\s^2_1+\s^2_3)~.\eqno(4.5)$$
The conformal structure, associated with the metric (4.5) inside the unit ball 
in ${\bf C}^2$, does not extend across the boundary since the coefficient at 
$\s^2_2$ decays {\it faster} than the coefficients at  $\s^2_1$ and  $\s^2_3$.
However, the conformal structure survives  in the {\it two-dimensional} (2d) 
subspace of $S^3$, which is annihilated by $\s_2$, because it is protected by 
the K\"ahler nature of the metric (4.5).

The Bergmann metric of the symmetric space $SU(2,1)/U(1)\times SU(2)$ can be 
identified with the Zamolodchikov metric of certain 2d, N=2 SCFT (sect.~5), 
which may serve as the UV fixed point for the RG flow. This equally aplies
to any Wolf space, while the associated 2d, N=2 SCFT to be defined via the
Kazama-Suzuki construction, in fact, possesses 2d, N=4 superconformal 
symmetry, albeit with the quadratically generated (Bershadsky-Knizhnik) 
algebra \cite{gket}. The formal central charge 
of the 2d SCFT, associated with $SU(2,1)/U(1)\times SU(2)$, is given by 
$c=3(3p+1)/(p+3)$ \cite{gket}. Note that $c\to 9$ when $\abs{p}\to\infty$.

The hypermultiplet moduli space, arising in the type-IIA superstring 
compactification on Calabi-Yau threefolds, also obeys eq.~(4.1), while in the
tree approximation (without quantum corrections) it is known to be described 
by the quaternionic manifold $Q_{\rm tree}$ similar to 
$SU(2,1)/U(1)\times SU(2)$ \cite{fst}. However, unlike the 
$SU(2,1)/U(1)\times SU(2)$, the $Q_{\rm tree}$ has the Heisenberg (isotropy) 
symmetry group instead of $SU(2)$, and thus belongs to the Bianchi II type.

\section{Weyl self-duality and RG flow}

The non-homogeneous solutions to eq.~(4.1), which can be interpreted as the RG
 flow, are supposed to share the basic features of the latter, namely,
\begin{itemize} 
\item they have to obey the first-order differential equations, 
\item there should be a well-defined RG flow parameter. 
\end{itemize}

The second condition is most naturally ensured by the $SU(2)$ isometry of the 
four-dimensional Zamolodchikov metric because the non-degenerate action of 
this isometry leads to the well-defined three-dimensional orbits that can be 
parametrized by the `radial'  coordinate $(t)$ to be identified with the RG 
parameter. In the context of N=2 supersymmetry, the $SU(2)$ isometry  has its 
origin in the unbroken R-symmetry $SU(2)$. The (Weyl) ASD equations on the 
metric take the form of a first-order system of Ordinary Differential 
Equations (ODE), so that the first condition above is automatic. In other 
words, in the context of the holographic duality, we should add the $SU(2)$ 
symmetry to the general requirements of eq.~(4.1), all dictated by the 
four-dimensional local N=2 supersymmetry alone. 

We are thus led to a study of the $SU(2)$-invariant deformations of the 
Bergmann metric (4.5) subject to the differential constraints (4.1). This 
well-defined mathematical problem was addressed by Tod \cite{tod} and Hitchin 
\cite{hit}. A generic $SU(2)$ invariant metric in the Bianchi IX formalism 
reads
$$ ds^2=w_1w_2w_3dt^2+\fracmm{w_2w_3}{w_1}\s^2_1+\fracmm{w_3w_1}{w_2}\s^2_2+  
\fracmm{w_1w_2}{w_3}\s^2_3~,\eqno(5.1)$$
where we have taken it in the diagonal form with respect to the $\s_i$ of
eq.~(4.3), without loss of generality. As
was demonstrated, e.g., in ref.~\cite{tod}, the Weyl ASD conditions (4.1)
applied to the {\it Ansatz} (5.1) result in the classical Halphen system of 
ODE \cite{halphen},
$$\eqalign{
 \dt{A}_1~=~&-A_2A_3 +A_1(A_2+A_3) ~,\cr
 \dt{A}_2~=~&-A_3A_1 +A_2(A_3+A_1) ~,\cr
 \dt{A}_3~=~&-A_1A_2 +A_3(A_1+A_2) ~,\cr}\eqno(5.2)$$
where the dots denote differentiation with respect to $t$, and the functions 
$A_i$,  $i=1,2,3$, are defined by the auxiliary system of ODE,
$$\eqalign{
   \dt{w}_1~=~&-w_2w_3 +w_1(A_2+A_3)~,\cr
   \dt{w}_2~=~&-w_3w_1 +w_2(A_3+A_1)~,\cr
   \dt{w}_3~=~&-w_1w_2 +w_3(A_1+A_2)~.\cr}\eqno(5.3)$$

The Bergmann metric corresponds to the dual 2d SCFT since all its $A_i$ 
vanish, as expected: this follows from a comparison of eqs.~(4.5), (5.1) and
(5.3). Being considered as the SCFT Zamolodchikov metric, the metric (4.5) is 
non-trivial (i.e. non-flat), even though all $A_i=0$.

Being of the form
$$ \dt{\l}_i=C_{ijk}\l_j\l_k~,\eqno(5.4)$$
the ODE system (5.2) is the particular case of the RG flow equations 
(2.4) in the dual 2d, N=2 QFT, whose coefficients $C_{ijk}$ represent the 
{\it normalized} (and universal) OPE coeffcients of the underlying 2d, N=2 
SCFT at the UV fixed point of the 2d, N=2 QFT. The Zamolodchikov $c$-function 
defined by
$$ c(\l)= c - \fracm{1}{3}C_{ijk}\l_i\l_j\l_k \eqno(5.5)$$
 satisfies the Zamolodchikov condition, 
$$ \dt{c}\,=-C_{ijk}\l_i\l_j\dt{\l}_k = 
 -\sum_i(\dt{\l}_i)^2\leq 0~,\eqno(5.6)$$
for any choice of the totally symmetric coefficients $C_{ijk}$ in eq.~(5.4).

The second (Einstein) condition in eq.~(4.1) can be easily satisfied by 
conformal rescaling of a solution to the (Weyl) ASD metric provided by the
ODE systems (5.2) and (5.3), because any local Weyl transformation does not 
affect the vanishing Weyl tensor (see sect.~6 for details). Having obtained 
an explicit solution to the Halphen system (5.2), it may be substituted into
 eq.~(5.3). To solve eq.~(5.3), it is convenient to change variables as 
\cite{tod}
$$ \eqalign{
 w_1~=~&\fracmm{\O_1\dt{x}}{\sqrt{x(1-x)}}~,\cr
 w_2~=~&\fracmm{\O_2\dt{x}}{\sqrt{x^2(1-x)}}~,\cr
 w_3~=~&\fracmm{\O_3\dt{x}}{\sqrt{x(1-x)^2}}~~,\cr}\eqno(5.7)$$
where the new variables $\O_i(x)$, $i=1,2,3$, are constrained by the algebraic
condition 
$$\O_2^2+\O_3^2-\O^2_1=\fracm{1}{4} \eqno(5.8) $$
that reduces the number of the newly introduced functions in eq.~(5.7) from 
four to three, as it should.

Equations (5.3) in terms of the new variables take the form \cite{tod,hit}
$$\eqalign{
 \O_1'~=~&-\fracmm{\O_2\O_3}{x(1-x)}~,\cr
 \O_2'~=~&-\fracmm{\O_3\O_1}{x}~,\cr
 \O_3'~=~&-\fracmm{\O_1\O_2}{1-x}~,\cr}\eqno(5.9)$$
where the primes denote differentiation with respect to $x$. It is not
difficult to verify that the algebaric constraint (5.8) is preserved under 
the flow (5.9), so that the transformation (5.7) is fully consistent. In
terms of the new variables $(x,\O_i)$, the Einstein condition of eq.~(4.1) on 
the metric (5.1), having the form
$$ ds^2=e^{2u}\left[ \fracmm{dx^2}{x(1-x)}+\fracmm{\s^2_1}{\O^2_1}
+\fracmm{(1-x)\s^2_2}{\O^2_2}+\fracmm{x\s^2_3}{\O^2_3}\right]~,\eqno(5.10)$$
amounts to the algebraic relation \cite{tod}
$$ 96\k^2e^{2u}=\fracmm{8x\O^2_1\O^2_2\O^2_3+2\O_1\O_2\O_3(x(\O_1^2+\O_2^2)-
(1-4\O^2_3)(\O^2_2-(1-x)\O ^2_1))}{(x\O_1\O_2+2\O_3(\O_2^2-(1-x)\O^2_1))^2}~.
\eqno(5.11)$$

Having interpreted the ODE system (5.4) as the RG flow equations in the 
(non-conformal) dual 2d, N=2 QFT originating from 2d, N=4 SCFT, the crucial 
question is, of course, about the behaviour of the RG flow in the IR limit,
$x\to 1^-$. The holographic interpretation of an exact metric solution to 
eq.~(4.1) requires it to be regular (or complete) in the bulk, so that all 
of its pole singularities  (in a particular parametrization) have to be 
removable by coordinate transformations. The existence of an IR fixed point 
implies that the regular metric should have the asymptotical behaviour 
similar to that of the Bermann metric, i.e. the metric coefficient at $\s^2_2$
 in eq.~(5.1) should decay faster than the others. This would mean that the 
boundary annihilated by $\s_2$ is two-dimensional, while it has a conformal
 structure and a conformal anomaly. The K\"ahler structure in the bulk is 
going to be extendable to the boundary, which implies that the 2d CFT on the 
boundary should be N=2 supersymmetric, at least. The explicit regular metric 
solutions (sect.~6) confirm these expectations.

\section{Painlev\'e VI equation and complete solution}

The ODE system (5.2) has a long history \cite{abc}. Perhaps, its most
natural (manifestly integrable) derivation is provided via a reduction of 
the $SL(2,{\bf C})$ anti-self-dual Yang-Mills equations from four Euclidean 
dimensions to one  \cite{int}. A classification of all possible reductions is 
known in terms of the so-called {\it Painlev\'e} groups that give rise to six 
different types of integrable Painlev\'e equations \cite{int}. It remains to 
identify those of them that lay behind the Weyl-ASD (quaternionic-K\"ahler) 
geometry with $SU(2)$ symmetry. There are only two natural (or nilpotent, in 
the terminology of ref.~\cite{int}) types (III and VI) that give rise to a 
single non-linear integrable equation. In the geometrical terms, it is the 
Painlev\'e III equation that lays behind the four-dimensional K\"ahler spaces 
with vanishing scalar curvature \cite{pp}, whereas the Painlev\'e VI equation 
is known to be behind the Weyl-ASD geometries having the $SU(2)$ symmetry 
\cite{tod,hit,m3}. A generic Painlev\'e VI equation has four real parameters 
\cite{int}, but they are all fixed by the quaternionic-K\"ahler property 
\cite{tod,hit}. This results in the following particular Painlev\'e VI 
equation:
$$\eqalign{
y''~=~&\fracmm{1}{2}\left( \fracmm{1}{y} 
+\fracmm{1}{y-1}+\fracmm{1}{y-x}\right)
(y')^2- \left( \fracmm{1}{x} +\fracmm{1}{x-1}+\fracmm{1}{y-x}\right)y'\cr
&~  +\fracmm{y(y-1)(y-x)}{x^2(x-1)^2}\left[
\fracmm{1}{8} -\fracmm{x}{8y^2}+\fracmm{x-1}{8(y-1)^2}
+\fracmm{3x(x-1)}{8(y-x)^2}\right] ~,\cr}\eqno(6.1)$$
where $y=y(x)$, and the primes denote differentiation with respect to $x$.

The equivalence between eqs.~(5.2) and (6.1) via eq.~(5.9) is 
well known to mathematicians \cite{tod,hit,m3}. Explicitly, in the Einstein 
case, it reads
$$\eqalign{
\O^2_1~=~&\fracmm{(y-x)^2y(y-1)}{x(1-x)}\left(v-\fracmm{1}{2(y-1)}\right)\left(
v-\fracmm{1}{2y}\right)~,\cr
\O^2_2~=~&\fracmm{(y-x)y^2(y-1)}{x}\left(v-\fracmm{1}{2(y-x)}\right)\left(
v-\fracmm{1}{2(y-1)}\right)~,\cr
\O^2_3~=~&\fracmm{(y-x)y(y-1)^2}{(1-x)}\left(v-\fracmm{1}{2y}\right)\left(
v-\fracmm{1}{2(y-x)}\right)~,\cr}\eqno(6.2)$$
where the auxiliary variable $v$ is defined by the equation
$$ y'=\fracmm{y(y-1)(y-x)}{x(x-1)}\left(2v-\fracmm{1}{2y}-\fracmm{1}{2(y-1)}
+\fracmm{1}{2(y-x)}\right)~.\eqno(6.3)$$

An exact solution to the Painlev\'e equation (6.1), which  leads to a 
{\it complete} (regular) metric, is known to be unique, while it can be 
expressed in terms of the standard theta-functions $\vq_{\a}(z|\t)$ where 
$\a=1,2,3,4$. We use the standard definitions and notation for the theta 
functions \cite{theta} --- see Appendix. In order to write down the relevant
solution to eq.~(6.1), the theta-function arguments should be related by
 $z=\fracm{1}{2}(\t -k)$, where $k$ is considered to be an arbitrary (real and
positive) parameter. The relation to the $x$-variable of eq.~(6.1) is given by
$x=\vq^4_3(0)/\vq^4_4(0)$, where the value of $z$ is explicitly indicated, as
usual. One finds \cite{dub,hit}
$$\eqalign{
y(x)~=~& \fracmm{\vq_1'''(0)}{3\p^2\vq^4_4(0)\vq_1'(0)}+\fracmm{1}{3}\left[
1+\fracmm{\vq_3^4(0)}{\vq^4_4(0)}\right] \cr
 &~ +\fracmm{\vq_1'''(z)\vq_1(z)-2\vq_1''(z)\vq_1'(z)+
2\p i(\vq_1''(z)\vq_1(z)-\vq_1'{}^2(z))}{2\p^2\vq_4^4(0)\vq_1(z)(\vq_1'(z)+
\p i\vq_1(z))}~~.\cr}\eqno(6.4)$$ 

The parameter $k>0$ describes the monodromy of the solution (6.4) around 
its essential singularities (branch points) $x=0,1,\infty$. This (non-abelian)
 monodromy is generated by the matrices (with the eigenvalues $\pm i$)
 \cite{hit}
$$ M_1=\left( \begin{array}{cc} 0 & i \\ i & 0\end{array} \right)~,\quad
  M_2=\left( \begin{array}{cc} 0 & i^{1-k} \\ 
i^{1+k} & 0\end{array} \right)~,\quad
M_3=\left( \begin{array}{cc} 0 & i^{-k} \\ 
-i^{k} & 0\end{array} \right)~.\eqno(6.5)$$
The explicit (equivalent) form of an exact solution to the metric 
coefficients $w_i$ in eq.~(5.3) was derived in ref.~\cite{bk}, in terms of the
 theta functions with characteristics, by the use of the fundamental 
Schlesinger system and the isomonodromic deformation techniques. 

The function (6.4) is meromorphic outside $x=0,1,\infty$, with the simple poles
at $\bar{x}_1,\bar{x}_2,\ldots$, where $\bar{x}_n\in (x_n,x_{n+1})$ and 
$x_n=x(ik/(2n-1))$ for each positive integer $n$. Accordingly, the metric is 
well-defined (complete) for $x\in (\bar{x}_n,x_{n+1}]$, i.e. in the unit ball 
with the origin at $x=x_{n+1}$ and the boundary at $x=\bar{x}_n$ \cite{hit}.
Near the boundary the metric (11) has the asymptotical behaviour 
$$\eqalign{
 ds^2~=~&\fracmm{dx^2}{(1-x)^2}+\fracmm{4}{(1-x)\cosh^2(\p k/2)}\s^2_1+
\fracmm{16}{(1-x)^2\sinh^2(\p k/2)\cosh^2(\p k/2)}\s^2_2\cr
&~ + \fracmm{4}{(1-x)\sinh^2(\p k/2)}\s^2_3+ ~~{\rm regular~terms}~.\cr}
\eqno(6.6)$$
As is clear from eq.~(6.6), the coefficient at $\s^2_2$ vanishes faster than
the coefficients at $\s^2_1$ and $\s^2_3$ when approaching the boundary,
$x\to 1^-$, similarly to eq.~(4.5), so that there is the natural conformal 
structure,
$$ \sinh^2(\p k/2)\s^2_1 +\cosh^2(\p k/2)\s^2_3~, \eqno(6.7)$$
on the two-dimensional boundary annihilated by $\s_2$ \cite{tod,hit}. The only 
relevant parameter $\tanh^2(\p k/2)$ in eq.~(6.7) represents the central 
charge (the conformal anomaly, or the critical exponent) of the 
two-dimensional superconformal field theory on the boundary. In the interior of
the ball we have the spectral flow, with the monotonically decreasing 
`effective' central charge (i.e. the $c$-function), in full accord with 
the c-theorem \cite{zam2}. The RG evolution ends at another (IR) fixed point 
where the solution (6.4) has a removable pole. This IR-fixed point thus may be
called a supersymmetric attractor.

\section{Conclusion} 

The proposed holographic duality gives the simple and natural description of 
the 2d RG-flow in 2d, N=2 QFT in terms of the effective (internal `gravity') 
NLSM in the background of N=2 supergravity in four spacetime dimensions. The
 local 4d, N=2 supersymmetry appears to be the sole source of the fundamental 
constraints on the NLSM (Zamolodchikov) metric. The regular $SU(2)$-invariant 
four-dimensional solutions to the Zamolodchikov metric are unique, being 
parametrized by the SCFT central charge describing the monodromy of the 
`master' solution to the underlying Painlev\'e VI equation. Our geometrical 
description of the RG flow by the quaternionic-K\"ahler geometry or eq.~(4.1) 
is manifestly covariant with respect to arbitrary reparametrizations of the 2d
 QFT coupling constants  --- {\it cf.} ref.~\cite{busso}.

Cecotti and Vafa \cite{cv} studied the integrable deformations of the 2d, N=2 
superconformal Landau-Ginzburg models by the most relevant operators (see 
ref.~\cite{fmvw} also). They identified the Zamolodchikov metric with the 
metric of the supersymmetric 
ground states, and found that it satisfies the classical Toda-like equations 
whose solutions are governed by the Painlev\'e III equation. The Cecotti-Vafa 
solutions are apparently associated with the K\"ahler metrics of vanishing 
scalar curvature \cite{pp} when the background gravity decouples, $\k=0$. In 
our explicit example, the RG flow in a 2d, N=2 QFT is described by the ODE 
system (5.2) whose coefficients are the universal (normalized) OPE coeffients 
of the underlying CFT at the UV-fixed point of the QFT. Unlike the 2d, 
N=2 supersymmetric RG flow solutions found by Cecotti and Vafa \cite{cv}, 
our RG flow has an IR fixed point and, therefore, it  can be interpreted 
as a domain-wall solution.

The constraints (4.1) do not seem to imply any quantization condition on the 
monodromy parameter $k$ since the regular metric solutions exist for any $k>0$,
 whereas the related central charge (or the critical exponent $k$ in
eq.~(6.7)) on the two-dimensional boundary is usually quantized in solvable
2d, N=2 SCFT like, e.g., the minimal N=2 superconformal models associated with 
compact (simply-laced) Lie groups. A resolution of this puzzle may be related 
to the {\it negative} curvature of the metrics. The ASD Einstein metrics of 
{\it positive} curvature take the similar form given by eqs.~(5.1) or (5.10), 
while they are known to be related to the so-called Poncelet $n$-polygons that
 give rise to the quantization condition $k=2/n$, where $n\in {\bf Z}$ 
\cite{hit2}. So, it seems that the absence of quantization may be explained by
 the non-compact nature of the Lie group $SU(2,1)$. Perhaps, the Tod-Hitchin 
metrics may also be interpreted as the kink-type solitons preserving some 
supersymmetry, i.e. as the BPS-type solutions in the context of 
higher-dimensional supergravity ({\it cf.} ref.~\cite{flow}). It would also be
 interesting to investigate their possible connections to matrix models, 
2d gravity and non-commutative geometry.

A supersymmetric version of the Randall-Sundrum scenario \cite{rs} with a
gravity localized near the wall under the exponential suppression, recently 
attracted much attention, partly because it appears to be impossible without 
the use of hypermultiplets \cite{srs,srs1,srs2}. Though we didn't discuss here
 any solutions to the spacetime metric $g_{\m\n}$, the existence of the exact 
domain wall solutions to the RG flow associated with the NLSM in eq.~(2.9) may
 be related to the domain-walls in spacetime via the Einstein equations. In
particular, the need of IR fixed points for the existence of regular and 
supersymmetric Randall-Sundrum type domain-wall solutions in spacetime was 
emphasized in ref.~\cite{srs}.

\section*{Acknowledgements}

I would like to thank Ioannis Bakas and Elias Kiritsis, the Organizers of the
EURESCO Conference `Quantum Fields and Strings', for a kind hospitality 
extended to me in Crete, where a part of this work was done. I am also grateful
 to all participants of the Conference for stimulating atmosphere and 
illuminating discussions.

\section*{Appendix: Basic facts about theta-functions}

The first theta-function $\vq_1(z|\t)$  is defined by the series \cite{theta}
$$\eqalign{
 \vq_1(z) \equiv  \vq_1(z|\t) & = 
-i\sum^{+\infty}_{n=-\infty}(-1)^n\exp i \left\{
\left( n +\fracm{1}{2}\right)^2\p\t + (2n+1)z\right\} \cr
 & = 2\sum^{+\infty}_{n=0}(-1)^n q^{(n+1/2)^2}\sin(2n+1)z~,\quad
q=e^{i\p\t}~, \cr} \eqno(A.1)$$
where $\t$ is regarded as the fundamental complex parameter, whose imaginary
part must be positive, $q$ is called the nome of the theta-function,
$\abs{q}<1$, and $z$ is the complex variable. The other theta-functions are
defined by \cite{theta}
$$\eqalign{
\vq_2(z|\t) & = ~\vq_1(z+\fracm{1}{2}\p)|\t) =
 \sum^{+\infty}_{n=-\infty} q^{(n+1/2)^2}e^{i(2n+1)z} \cr
& =  2\sum^{+\infty}_{n=0} q^{(n+1/2)^2}\cos(2n+1)z~,\cr}\eqno(A.2)$$
$$\eqalign{
\vq_3(z|\t) & = ~\vq_4(z+\fracm{1}{2}\p)|\t) =
 \sum^{+\infty}_{n=-\infty} q^{n^2}e^{2inz} \cr
& = 1+ 2\sum^{+\infty}_{n=1} q^{n^2}\cos 2nz~,\cr} \eqno(A.3)$$
and
$$ \vq_4(z|\t) = \sum^{+\infty}_{n=-\infty} (-1)^n q^{n^2}e^{2inz} =
1+ 2\sum^{+\infty}_{n=1} (-1)^n q^{n^2}\cos 2nz~.\eqno(A.4)$$

The identities \cite{theta}
$$ \vq^4_3(0) = \vq^4_2(0)+\vq^4_4(0)~,\eqno(A.5)$$
$$ \vq_1'(0) = \vq_2(0)\vq_3(0)\vq_4(0)~,\eqno(A.6)$$
and
$$ \fracmm{\vq_1'''(0)}{\vq_1'(0)}= 
\fracmm{\vq_2''(0)}{\vq_2(0)}+\fracmm{\vq_3''(0)}{\vq_3(0)}+
\fracmm{\vq_4''(0)}{\vq_4(0)}~~~,\eqno(A.7)$$
where the primes denote differentiation with respect to $z$, may be used to
 rewrite eq.~(6.4) to other equivalent forms (cf.~\cite{hit,dub,bk}).

\end{document}